\documentclass[prd,nofootinbib,amsfonts,notitlepage,showpacs]{revtex4-1}
\usepackage{graphicx,bm,amsmath,color}
\usepackage[latin1]{inputenc}
\usepackage{soul,hyperref,verbatim}

\newcommand{\be}{\begin{equation}}
\newcommand{\ee}{\end{equation}}
\newcommand{\beq}{\begin{eqnarray}}
\newcommand{\eeq}{\end{eqnarray}}
\newcommand{\comments}[1]{}

\newcommand{\sqamp}{\langle |\mathcal{M}|^2\rangle}

\begin{document}

\title{Uncertainties in Constraints from Pair Production on Superluminal Neutrinos}
\author{J.M. Carmona}
\email{jcarmona, cortes@unizar.es, dg.mazon@gmail.com}
\affiliation{Departamento de F\'{\i}sica Te\'orica,
Universidad de Zaragoza, Zaragoza 50009, Spain}
\author{J.L. Cort\'es}
\email{jcarmona, cortes@unizar.es, dg.mazon@gmail.com}
\affiliation{Departamento de F\'{\i}sica Te\'orica,
Universidad de Zaragoza, Zaragoza 50009, Spain}
\author{D. Maz\'on}
\email{jcarmona, cortes@unizar.es, dg.mazon@gmail.com}
\affiliation{Departamento de F\'{\i}sica Te\'orica,
Universidad de Zaragoza, Zaragoza 50009, Spain}

\begin{abstract}
The use of the vacuum lepton pair production process ($\nu  \to \nu \, e^- \, e^+ $), a viable reaction for superluminal neutrinos, to put constraints on Lorentz violations requires a dynamical framework. Different choices of dynamical matrix elements and modified dispersion relations for neutrinos, leading to numerical factors differing by one order of magnitude in the results for the pair production decay width, are used to show the uncertainties on these constraints.
\end{abstract}


\maketitle

\section{Introduction}
Lorentz invariance is one of the pillars of the current theories of  fundamental physics. It implements the physical equivalence of inertial
observers in absence of gravity, together with the presence of an invariant scale, the speed of light in vacuum. In particular, Lorentz
symmetry is an important ingredient of the Standard Model (SM) of elementary particles. However, the logical consistency of the latter
does not rely on the former, contrary to what happens to unitarity or the absence of gauge anomalies which are necessary requirements
for a quantum theory like the SM. This permits investigating departures from Lorentz invariance in the framework of a quantum field
theory (QFT), and therefore, testing Lorentz invariance as a symmetry of the SM. Although much progress has been made in this direction,
the non-compactness nature of the Lorentz group, the fact that the boost parameter can be arbitrarily large, implies that the unexplored region of this group will always be infinitely great, unlike the rotation group which is compact and can be checked for every angle. Apart from the own interest of testing such a fundamental symmetry, there are motivations to study deviations from Lorentz invariance coming from both the theoretical and the phenomenological sides. On the one hand, some indications coming mostly from theoretical investigations of the quantum-gravity problem have suggested that Lorentz symmetry could be either violated or deformed at high enough energies (see~\cite{BW,ST,LQG,RG,AM,Girelli:2004md,Smolin:2008hd}). On the other hand, some observations might be interpreted to be incompatible with Lorentz invariance, such as a possible violation of the GZK cut-off~\cite{GZK}, anomalies in the propagation of very high energy gamma rays in intergalactic space~\cite{meyer,piran}, or, more recently, the observation of  apparent superluminal neutrinos by the OPERA collaboration~\cite{Adam:2011zb} following a proposal~\cite{Ellis:2008fc} to look for signals of Lorentz invariance violation in neutrino propagation.

In this work, we present an analysis of the consequences of such departures from exact Lorentz invariance in the decay of superluminal neutrinos. In particular, we shall focus on charged-lepton pair emission by superluminal neutrinos ($\nu _i \to \nu _i \; e^- \; e^+ $) which is a forbidden process in a Poincar\'e invariant theory because of the charged lepton's masses. The idea that neutrinos can travel faster than light is not recent, and it actually goes back to the eighties (see~\cite{Chodos1984cy} and~\cite{Giannetto:1986rm}).  The difficulty in the detection of neutrinos (owing to the weakness of their interactions with matter) entails some of their features remain still unknown, making neutrinos somehow mysterious particles. Since neutrino oscillations suggest that they are not massless particles, it is natural to expect the existence of right-handed neutrinos which are transparent to the electroweak and strong interactions of the SM (they are singlets under $U_Y(1)\times SU_L(2)\times SU_C(3)$ transformations). This special property has been used in the large-extra-dimensions scenario to speculate about the possibility  that neutrinos could not be confined to our three-brane, but they are indeed exploring  the extra dimensions~\cite{Ark}. Because of this, if the departures from Lorentz invariance had their origin in these additional dimensions, it would be expected that neutrinos were the unique particles capable of directly feeling such deviations from exact Lorentz symmetry (perhaps together with an hypothetical graviton which would also be transparent to the gauge interactions of the SM). The particular quantum numbers of neutrinos in the SM make also possible to write a gauge invariant but Lorentz violating Lagrangian which gives rise to effects of Lorentz invariance violation which are different for neutrinos than for the rest of the SM particles.
Other works have considered neutrinos as the only candidates for Lorentz violating particles in an effective field theory (EFT) framework because in this context, and in presence of interactions, different limiting speeds of different particles at high energies are driven by the renormalization group flow towards a universal speed at low energies (although strong assumptions have to be made in order that the flow be sufficiently fast), the vacuum speed of light (see~\cite{Don}). The weaker the interaction, the weaker the running and therefore the difference of limiting speeds is the greatest when the interactions are the weakest. This suggests that the limiting speed of neutrinos at low energies may be different from the speed of the rest of particles  when Lorentz invariance has been broken at high energies.

In this paper, we shall parameterize the deviations from exact Lorentz invariance by a modified dispersion relation for free neutrinos. The consequence of such modification is twofold: on the one hand, there is a new dependence of the velocity on the energy which makes the time of flight different from the one of a special-relativistic theory. On the other hand, the kinematics of particle processes involving neutrinos changes. In fact the dynamics of these processes is also changed because of the different choices of a dynamical matrix element compatible with the modified dispersion relation one can make. Previous works have focused on getting qualitative or quantitative results for the decay width $\nu_i \to \nu_i \, e^- \, e^+ $ or $\nu_i \to \nu_i \, \nu_j  \, \bar{\nu_j} $ \cite{Mat,Cohen:2011hx,Carmona:2011zg,Maccione:2011fr,Ciuffoli:2011aa}, without discussing the different alternatives one has for the matrix element, or even without specifying the matrix element they are using in their calculation (however, see~\cite{Bez}). The main \emph{purpose} of the present work is to show how this can affect the results for this process. One could think that a criterion to select a matrix element would be that it could be deduced from a local EFT. However, there exist limitations to this framework. First, one should restrict dispersion relations to analytical expressions so that a momentum power expansion be possible in the EFT. Second, the most likely origin of such deviations from Poincar\'e symmetry arises in attempts to reconcile general relativity and quantum theory, through residual effects (that is, effects that are present when the classical curvature of space-time can be neglected) in the structure of space-time or/and momentum space which modify its classical, Minkowskian nature~\cite{CNC,Kow}. If this were the case, it would not be surprising that this led to non-local effects at low energies since the semi-classical physics of black holes tells us that the fundamental degrees of freedom of gravity cannot be described by a \emph{local} QFT, regardless of whether the latter has a non-trivial fixed point or not, because the densities of states of both theories do not match~\cite{Susskind:1994vu,Shomer}.
Furthermore, doubts on the validity of an EFT description of the low energy limit of quantum gravity come from the difficulties to incorporate the necessary cancellations of contributions to the vacuum energy (cosmological constant problem).

Throughout this work, we shall assume: i) Rotational invariance is preserved. ii) Energy and momentum are conserved in the conventional, additive way. iii) The relevant propagation speed of the neutrinos is the group velocity of their wave packets. iv) Charged leptons are (or can be approximated by) special-relativistic particles. v) Indirect bounds on neutrino masses based on neutrino oscillations and cosmological observations are still valid, in such a way that neutrinos masses can be neglected in the process.

This work has the following structure. In the next section~(\ref{SecPair}) we shall show the general procedure to obtain the decay widths, and we will compute them for four different choices of the matrix element and a general dispersion relation. Section~\ref{SecChoices} is devoted to discuss the properties of the previous matrix elements and their physical interpretation. In Section~\ref{SecEnergy} we shall specify dispersion relations to give definite analytical results for the decay widths and the rates of energy loss due to the pair emission. We shall also provide numerical results for these expressions. Section~\ref{SecTimes} is dedicated to the study of the time of flight of neutrinos in the presence of the aforementioned loss of energy. The following section~(\ref{SecConsistency}) is devoted to analyze the consistency of the superluminal speed interpretation of the observations reported by the OPERA collaboration. Our conclusions and remarks will close this work in Sec.~\ref{SecConcluiding}.

\section{Pair production decay width}
\label{SecPair}
Let us consider the process $\nu(p) \to \nu(p_1) \, e^{-}(p_2) \, e^{+}(p_3)$ induced by the production of a virtual $Z^0$ and subsequent decay into an $e^{-}e^{+}$ pair. This is a kinematically forbidden process in special relativity (SR) which becomes allowed for superluminal neutrinos above a certain energy threshold. We are going to consider a very high energy  neutrino so that all three particles in the final state are contained within a small cone around the direction of the momentum of the neutrino in the initial state.  The decay width of the process is given by
\be
\Gamma =\frac{1}{2E} \left[\prod_{i=1}^3 \int \frac{d^3 {\vec p}_i}{2 E_i (2\pi)^3}\right] (2\pi)^4 \delta(E-\sum_i E_i) \delta^3 ({\vec p}-\sum_i {\vec p}_i) \, \sqamp \,,
\label{Gamma(pp)}
\ee
where $\sqamp$ is the squared amplitude averaged over initial spin states and summed over final spin states,
\be
\sqamp = A^{\mu\nu}(p, p_1) p_{2\mu} p_{3\nu}\,,
\label{def(A)}
\ee
and we are using the factorization of the matrix element into a factor depending on the neutrino variables and a factor depending on the electron-positron momenta. The second factor is the standard SR Dirac trace which gives the dependence on the four-momenta $p_2$ and $p_3$. The coefficients $A^{\mu\nu}$ depending on the neutrino momenta $p$, $p_1$ contain all the corrections to SR in the dynamical matrix element.

The decay width can be written as
\be
\Gamma = \frac{1}{2E} \int \frac{d^3 \vec{p}_1}{2E_1 (2\pi)^3} A^{\mu\nu}(p, p_1) B_{\mu\nu}(p-p_1)
\label{Gammacov}
\ee
where
\be
B_{\mu\nu}(k) = \int \frac{d^3\vec{p}_2}{2E_2 (2\pi)^3} \int \frac{d^3\vec{p}_3}{2E_3 (2\pi)^3} \, p_{2\mu} p_{3\nu} (2\pi)^4 \delta^4(k-p_2-p_3) \,.
\ee
In the approximation where one neglects any modification to SR kinematics for the electron and positron,
\be
B_{\mu\nu}(k) = B_1(k^2) \eta_{\mu\nu} k^2 + B_2(k^2) k_\mu k_\nu \,,
\ee
where $k^2 = k_0^2 - {\vec k}^2$, and
\begin{align}
B_1(k^2) &= \frac{1}{96\,\pi} \left(1 - \frac{4m_e^2}{k^2}\right)^{3/2}\,
\theta(k^2 - 4m_e^2)\, \theta(k_0) \nonumber \\
B_2(k^2) &= \frac{1}{96\,\pi} \left(2 + \frac{4m_e^2}{k^2}\right)
\left(1 - \frac{4m_e^2}{k^2}\right)^{1/2} \theta(k^2 - 4m_e^2)\, \theta(k_0) \,,
\end{align}
where $\theta$ is the Heaviside step function. On the other hand, one has
\be
k^2 = (p-p_1)^2  = (E-E_1)^2 - (|\vec{p}|-|\vec{p}_1|)^2 - 2 |\vec{p}| |\vec{p}_1| (1-\cos\theta_1)
\label{k2}
\ee
where $\theta_1$ is the angle between the neutrino momenta $\vec{p}$, $\vec{p}_1$. The minimum value of $k^2$ ($4m_e^2$) corresponds to a maximum value of $(1-\cos\theta_1)$
\be
(1-\cos\theta_1)_{+} = \frac{(E-E_1)^2 - (|\vec{p}|-|\vec{p}_1|)^2 - 4 m_e^2}{2 |\vec{p}| |\vec{p}_1|}\,.
\ee
If we parameterize the energy-momentum relation for neutrinos as\footnote{We are neglecting neutrino masses in all the discussion.}
\be
E = |\vec{p}| \left(1 + \epsilon(|\vec{p}|)\right)
\label{E}
\ee
then we have
\be
(E-E_1)^2 - (|\vec{p}|-|\vec{p}_1|)^2 = \left[(E+|\vec{p}|) - (E_1+|\vec{p}_1|)\right] \left[|\vec{p}| \epsilon(|\vec{p}|) - |\vec{p}_1| \epsilon(|\vec{p}_1|)\right]
\ee
and one has a negative result for the maximum value of $(1-\cos\theta_1)$, i.e., pair production is not allowed, in the case of SR kinematics ($\epsilon(|\vec{p}|)=0$).  The threshold momentum $|\vec{p}_{th}|$ for pair production verifies:
\begin{equation}
|\vec{p}_{th}|^2\,\epsilon(|\vec{p}_{th}|)=2\, m_e^2 \,
\end{equation}
corresponding to the configuration in which the outgoing neutrino is at rest, and the electron and positron have the same momentum. We shall consider incoming neutrino momenta much greater than this threshold momentum, $|\vec{p}|\gg |\vec{p}_{th}|$, in such a way that one can safely neglect electron and positron masses. In addition, we shall assume that the deviation from SR kinematics is sufficiently small ($|\epsilon(|\vec{p}|)| \ll 1$) so that $|\vec{p}_1| (1-\cos\theta_1)_{+} \ll |\vec{p}|$. In this case, the change in the direction of the neutrino energy flux due to the emission of the $e^{-}e^{+}$ pair is very small (``collinear'' approximation).

We shall also restrict all the discussion to neutrino momenta such that
\be
\left[(E+|\vec{p}|) - (E_1+|\vec{p}_1|)\right] \left[|\vec{p}| \epsilon(|\vec{p}|) - |\vec{p}_1| \epsilon(|\vec{p}_1|)\right] \ll M_Z^2 \,.
\ee
In this case we can use the point like approximation for the fermion interaction in the pair production process and then the momentum dependence in the coefficients $A^{\mu\nu}(p, p_1)$, which parametrize all our ignorance on the dynamics, comes from the modified neutrino Dirac trace. Assuming a linear dependence on the components of each momentum in the amplitude then one has
a general form for the angular dependence
\be
\begin{split}
A^{\mu\nu}(p, p_1) B_{\mu\nu}(p-p_1) &= \frac{G_F^2 |\vec{p}|^4}{12 \pi} \left[(1-2s_W^2)^2 + (2s_W^2)^2 \right] \, \left[F_0(x_1) + F_1(x_1) (1-\cos\theta_1) + F_2(x_1) (1-\cos\theta_1)^2\right] \\
&\quad\times
\theta \left( (1-\cos\theta_1)_{+} - (1-\cos\theta_1)\right)\,\theta \left(1-x_1+\epsilon(|\vec{p}|) - x_1 \epsilon(x_1|\vec{p}|)\right) \, ,
\label{def(F)}
\end{split}
\ee
when one uses the momentum fraction $x_1 = |\vec{p}_1|/|\vec{p}|$ for the neutrino after pair production,
and where $G_F$ is the Fermi constant and $s_W$ is the sine of the Weinberg angle.
The determination of the coefficients $F_n(x_1)$ of the three angular terms requires a definite dynamical framework.
The angular integral on the neutrino momenta can be made\footnote{The angular integral in $(1-\cos\theta_1)$ goes from $0$ to $(1-\cos\theta_1)_{+} \,$,
except for those outgoing (very close to zero) momenta for which $2<(1-\cos\theta_1)_{+}$, in which case the angular integral goes from $0$ to $2\,$.} and the leading contribution for the decay width is
\be
\begin{split}
\Gamma = \frac {G_F^2 |\vec{p}|^5}{192 \, \pi^3} \left[(1-2s_W^2)^2 + (2s_W^2)^2 \right]\,
 \int _0 ^1  &\left[F_0(x_1) (1-\cos\theta_1)_{+}  + F_1(x_1) \frac{(1-\cos\theta_1)^2_{+}}{2}  +
F_2(x_1) \frac{(1-\cos\theta_1)^3_{+}}{3}\right] \\
& \times \theta \left(\epsilon(|\vec{p}|) - x_1 \epsilon(x_1|\vec{p}|)\right) x_1\, dx_1
\label{dGamma/dx1}
\end{split}
\ee
when one consistently retains only the dominant contribution in an expansion in powers of $\epsilon(|\vec{p}|)$.

\subsection{Dynamical matrix element}
Lacking a well defined dynamical framework incorporating superluminal neutrinos, we are going to consider four different simple choices for the dynamical matrix element as a way to illustrate the uncertainties in the evaluation of the pair production process. Naively, one would take the SR matrix element as a first approximation and one would expect that any other choice approaching the SR limit would lead to the same leading order result for the decay width but this is not the case. The reason is that in the SR limit and neglecting masses all four momenta $p$, $p_1$, $p_-$, $p_+$ are light-like and, due to energy-momentum conservation, proportional to each other so that any scalar product (and then the dynamical matrix element) vanishes.
Then the leading contribution to the dynamical matrix element comes from the first non-vanishing correction to the SR limit.

\subsubsection{First example}
An apparently natural choice for the dynamical matrix element corresponds to\footnote{We shall neglect masses in all the matrix elements.}
\be
A^{\mu\nu}(p,p_1) = 16 \, G_F^2 \left[(1-2s_W^2)^2 p_1^\mu p^\nu + (2s_W^2)^2 p^\mu p_1^\nu\right]
\label{A1}
\ee
which is the expression of the SR dynamical matrix element but with $p^0=|\vec{p}|(1+\epsilon(|\vec{p}|))$ and $p_1^0=|\vec{p}_1|(1+\epsilon(|\vec{p}_1|))$. In this case one has
\be
A^{\mu\nu}(p, p_1) B_{\mu\nu}(k) = \frac{G_F^2}{12 \, \pi} \, \left[(1-2s_W^2)^2 + (2s_W^2)^2 \right] \, \left[2 (p_1\cdot p) k^2  \,+\, 4 (p_1\cdot k) \, (p\cdot k)\right]\theta(k^2)\, \theta(k_0) \, .
\label{AB}
\ee

Using the products
\begin{align}
2 (p_1\cdot p) &= 2 \, \left[|\vec{p}| |\vec{p}_1| \left(\epsilon(|\vec{p}|) + \epsilon(|\vec{p}_1|)\right) + |\vec{p}| |\vec{p}_1| (1-\cos\theta_1)\right] \nonumber \\
k^2 &= 2 \, \left[(|\vec{p}|-|\vec{p}_1|) \left(|\vec{p}| \epsilon(|\vec{p}|) - |\vec{p}_1|
\epsilon(|\vec{p}_1|)\right) - |\vec{p}| |\vec{p}_1| (1-\cos\theta_1)\right] \nonumber \\
2 (p_1\cdot k) &= 2 \, \left[(|\vec{p}|-|\vec{p}_1|) |\vec{p}_1| \epsilon(|\vec{p}_1| +
|\vec{p}_1| \left(|\vec{p}| \epsilon(|\vec{p}|) - |\vec{p}_1| \epsilon(|\vec{p}_1|)\right) + |\vec{p}| |\vec{p}_1| (1-\cos\theta_1)\right] \nonumber \\
2 (p\cdot k) &= 2 \, \left[(|\vec{p}|-|\vec{p}_1|) |\vec{p}| \epsilon(|\vec{p}| +
|\vec{p}| \left(|\vec{p}| \epsilon(|\vec{p}|) - |\vec{p}_1| \epsilon(|\vec{p}_1|)\right) - |\vec{p}| |\vec{p}_1| (1-\cos\theta_1)\right]
\end{align}
we get
\begin{align}
F_0(x_1) &= 4 \left[x_1 (1-x_1)^2 \epsilon(|\vec{p}|) \, \epsilon(x_1|\vec{p}|) + 2 x_1 (1-x_1)  \left(\epsilon(|\vec{p}|) + \epsilon(x_1|\vec{p}|)\right) \left(\epsilon(|\vec{p}|) - x_1 \epsilon(x_1|\vec{p}|)\right) + x_1 \left(\epsilon(|\vec{p}|) - x_1 \epsilon(x_1|\vec{p}|)\right)^2\right], \nonumber \\
F_1(x_1) &= 4 \left[- x_1^2 \left(\epsilon(|\vec{p}|) + \epsilon(x_1|\vec{p}|)\right) + 3 x_1 (1-x_1) \left(\epsilon(|\vec{p}|) - x_1 \epsilon(x_1|\vec{p}|)\right)\right], \nonumber \\
F_2(x_1) &= - 8 x_1^2 \,.
\label{Fn}
\end{align}
When the angular integral is done one finds
\be
\begin{split}
\Gamma &= \frac {G_F^2 |\vec{p}|^5}{192 \, \pi^3}  \left[(1-2s_W^2)^2 + (2s_W^2)^2 \right] \, \int _0^1 dx_1 \,
\theta \left(\epsilon(|\vec{p}|) - x_1 \epsilon(x_1|\vec{p}|)\right) \,  \biggl[ 4 x_1 (1-x_1)^3 \epsilon(|\vec{p}|) \epsilon(x_1|\vec{p}|)
\left(\epsilon(|\vec{p}|) - x_1 \epsilon(x_1|\vec{p}|) \right)    \\
& \quad  + \,  6 x_1 (1-x_1)^2
\left(\epsilon(|\vec{p}|) + \epsilon(x_1|\vec{p}|)\right) \left(\epsilon(|\vec{p}|) - x_1 \epsilon(x_1|\vec{p}|)\right)^2 + \Bigl(4 x_1 (1-x_1)
+ \frac{10}{3} (1-x_1) ^3\Bigr) \left(\epsilon(|\vec{p}|) - x_1 \epsilon(x_1|\vec{p}|)\right)^3\biggl].
\end{split}
\label{Gamma}
\ee

If one is not interested in the angular dependence of the differential decay width, there is a simpler way to get the result in (\ref{Gamma}). Going back to (\ref{AB}), one can use the variable $k^2$ instead of the angle $\theta_1$ to express the products
\be
2 (p_1\cdot p) = p^2 + p_1^2 - k^2 {\hskip 1cm}
2 (p_1\cdot k) = p^2 - p_1^2 - k^2 {\hskip 1cm}
\,\, 2 (p\cdot k) = p^2 - p_1^2 + k^2
\ee
and
\be
A^{\mu\nu}(p, p_1) B_{\mu\nu}(k) = \frac{G_F^2}{12 \, \pi} \, \left[(1-2s_W^2)^2 + (2s_W^2)^2 \right] \, \left[(p^2 -p_1^2)^2 + (p^2 + p_1^2) k^2 - 2 (k^2)^2\right]\, \theta(k^2)\, \theta(k_0) \, .
\ee
The relation (\ref{k2}) allows replacing the angular integral by an integration over $k^2$, where the upper limit on $k^2$, corresponding to $\theta_1=0$, is given by
\be
k_+^2 = 2 |{\vec p}|^2 (1-x_1) \left[\epsilon(|{\vec p}|) - x_1 \epsilon(x_1|{\vec p}|)\right] \,.
\ee
This leads to:
\be
\Gamma = \frac {G_F^2}{192 \, \pi^3} \left[(1-2s_W^2)^2 + (2s_W^2)^2 \right] \, \frac{1}{2 |\vec{p}|}\,  \int _0 ^1 dx_1 \, \theta \left(k_+^2\right)  \left[(p^2 -p_1^2)^2 k_+^2 + \frac{1}{2} (p^2 + p_1^2) (k_+^2)^2  - \frac{2}{3} (k_+^2)^3\right].
\ee
Using
\be
p^2 = |\vec{p}|^2 (1 + 2 \epsilon(|\vec{p}|)) {\hskip 2cm}
p_1^2 = |\vec{p}_1|^2 (1 + 2 \epsilon(|\vec{p}_1|))
\ee
one recovers the result (\ref{Gamma}) for the decay width.

The choice (\ref{A1}) for the neutrino dependent factor in the dynamical matrix element cannot always be derived from a quantum field theoretical calculation. In fact, by considering a generic dispersion relation (arbitrary choice of $\epsilon(|\vec{p}|)$) one can have cases where Eq.~(\ref{AB}) and the expression of the decay width take negative values indicating an inconsistency of the \emph{ansatz} for the dynamical matrix element.

\subsubsection{Second example}
One could consider other alternatives to (\ref{A1}) for the matrix element. A very simple choice corresponds to consider a modified neutrino spinor satisfying a modified Dirac equation
\be
\left[\gamma^0 E(|\vec{p}|) - \vec{\gamma}\cdot \vec{p}\, (1+\epsilon(|\vec{p}|))\right] \tilde{u}(p) = 0\,.
\label{utilde}
\ee

This modified Dirac equation implies a modified dispersion relation $E(|\vec{p}|)=|\vec{p}| [1+\epsilon(|\vec{p}|)]$. With this modified Dirac neutrino spinors the matrix element can be calculated as in SR and the result for $A^{\mu\nu}$ is now
\be
\tilde{A}^{\mu\nu}(p,p_1) = 16 \, G_F^2 \left[(1-2s_W^2)^2 \tilde{p}_1^\mu \tilde{p}^\nu + (2s_W^2)^2 \tilde{p}^\mu \tilde{p}_1^\nu\right]
\label{A2}
\ee
with
\be
\tilde{p}^0 = |\vec{p}| [1+\epsilon(|\vec{p}|)] {\hskip 1cm}
\vec{\tilde{p}} = \vec{p}[1+\epsilon(|\vec{p}|)] \,.
\ee
This neutrino factor can be derived from a perturbative field theory calculation by considering the SR vertex for the  interaction and a modified free fermion action leading to the simple modification of the Dirac equation (\ref{utilde}).

With the choice (\ref{A2}) for the neutrino factor in the dynamical matrix element one has
\be
\begin{split}
\tilde{A}^{\mu\nu}(p, p_1) B_{\mu\nu}(k) &= 16 \, G_F^2 \left[(1-2s_W^2)^2 + (2s_W^2)^2 \right] \, \left[(\tilde{p}_1\cdot \tilde{p}) k^2 B_1(k^2) \,+\, (\tilde{p}_1\cdot k) \, (\tilde{p}\cdot k) B_2(k^2)\right] \\
&= \frac{G_F^2}{12 \pi} \left[(1-2s_W^2)^2 + (2s_W^2)^2 \right] \, \left[2 (\tilde{p}_1\cdot \tilde{p}) k^2 \,+\,4 (\tilde{p}_1\cdot k) \, (\tilde{p}\cdot k)\right]
\, \theta(k^2)\, \theta(k_0) \,.
\end{split}
\ee
In this case we have the products
\begin{align}
2 \tilde{p}_1\cdot \tilde{p} &= 2 |\vec{p}| |\vec{p}_1| (1-\cos\theta_1) \nonumber \\
k^2 &= 2 \, \left[(|\vec{p}|-|\vec{p}_1|) \left(|\vec{p}| \epsilon(|\vec{p}|) - |\vec{p}_1|
\epsilon(|\vec{p}_1|)\right) - |\vec{p}| |\vec{p}_1| (1-\cos\theta_1)\right] \nonumber \\
2 (\tilde{p}_1\cdot k) &= 2 |\vec{p}_1| \left(|\vec{p}| \epsilon(|\vec{p}|) - |\vec{p}_1|
\epsilon(|\vec{p}_1|)\right) + 2 |\vec{p}| |\vec{p}_1| (1-\cos\theta_1) \nonumber \\
2 (\tilde{p}\cdot k) &= 2 |\vec{p}| \left(|\vec{p}| \epsilon(|\vec{p}|) - |\vec{p}_1|
\epsilon(|\vec{p}_1|)\right) - 2 |\vec{p}| |\vec{p}_1| (1-\cos\theta_1)
\end{align}
and
\begin{align}
\tilde{F}_0(x_1) &= 4 x_1 \left(\epsilon(|\vec{p}|) - x_1 \epsilon(x_1|\vec{p}|)\right)^2\,, \nonumber \\
\tilde{F}_1(x_1) &= 8 x_1 (1-x_1) \left(\epsilon(|\vec{p}|) - x_1 \epsilon(x_1|\vec{p}|)\right), \nonumber \\
\tilde{F}_2(x_1) &= - 8 x_1^2 \,.
\label{Ftilden}
\end{align}
The decay width after the angular integration takes a different and simpler form than the decay width result of the first example (\ref{Gamma}):
\be
\tilde{\Gamma} = \frac {G_F^2 |\vec{p}|^5}{192 \, \pi^3} \left[(1-2s_W^2)^2 + (2s_W^2)^2 \right] \, \frac{4}{3} \, \int _0^1 dx_1 \, \theta \left(\epsilon(|\vec{p}|) - x_1 \epsilon( x_1|\vec{p}|)\right) \, (1 - x_1^3) \left(\epsilon(|\vec{p}|) - x_1 \epsilon( x_1|\vec{p}|)\right)^3 \,.
\label{Gammatilde}
\ee
Note that the result for the dynamical matrix element and then for the decay width is positive definite independently of the choice of the modified dispersion relation, so that the potential inconsistencies of the use of (\ref{A1}) for the neutrino factor are not present in this second example. This is a direct consequence of a field theoretical derivation of the dynamical matrix element.

\subsubsection{Third example}
One has a third simple choice for the dynamical matrix element with
\be
\bar{A}^{\mu\nu}(p,p_1) = 16 \, G_F^2 \left[(1-2s_W^2)^2 \bar{p}_1^\mu \bar{p}^\nu + (2s_W^2)^2 \bar{p}^\mu \bar{p}_1^\nu\right]
\label{A3}
\ee
with $\bar{p}^0 = |\vec{p}|$ and $\vec{\bar{p}} = \vec{p}$ which is just the SR dynamical matrix element.\footnote{Note that in the first choice of the dynamical matrix element one takes into account the modified expression for the energy at the level of the matrix element.} All one has to do is to replace in the first calculation everywhere $p$ by $\bar{p}$. Then one has
\be
\begin{split}
\bar{A}^{\mu\nu}(p, p_1) B_{\mu\nu}(k) &= 16 \, G_F^2 \left[(1-2s_W^2)^2 + (2s_W^2)^2 \right] \, \left[(\bar{p}_1\cdot \bar{p}) k^2 B_1(k^2) \,+\, (\bar{p}_1\cdot k) \, (\bar{p}\cdot k) B_2(k^2)\right] \\
&= \frac{G_F^2}{12 \pi} \left[(1-2s_W^2)^2 + (2s_W^2)^2 \right] \, \left[2 (\bar{p}_1\cdot \bar{p}) k^2 \,+\, 4 (\bar{p}_1\cdot k) \, (\bar{p}\cdot k)\right]
\, \theta(k^2)\, \theta(k_0) \,.
\end{split}
\ee
Due to the proportionality $\tilde{p}^\mu = \bar{p}^\mu (1+\epsilon(|\vec{p}|))$ it is obvious that the decay width in this third example coincides with the decay width in the second example up to corrections of order $\epsilon^4$.

\subsubsection{Fourth example}
In the particular case of a modified dispersion relation for the neutrino with a momentum independent choice for $\epsilon$, which corresponds to a momentum independent speed, it is possible to consider a modification of the interaction vertex fixed by gauge invariance from the modified free fermion action.
If we consider a Lagrangian
\be
\mathcal{L} = \bar{\nu}_L i \gamma^0 D_0 \nu _L - i\left(1+\frac{\eta_0}{2}\right) \bar{\nu}_L \, \vec{\gamma}\cdot \vec{D}\, \nu _L
\label{Leta0}
\ee
where $D_0$, $\vec{D}$ are covariant derivatives, then one has the simplest generalization of the relativistic Lagrangian for gauge interactions of massless fermions with a constant velocity $v = 1 + \eta_0/2$. In order to derive the dynamical matrix element all one has to do is to replace the Dirac $\gamma$ matrices $\gamma^\mu$ in the neutrino tensor by the modified $\gamma$ matrices $\hat{\gamma}^\mu$ where $\hat{\gamma}^0 = \gamma^0$ and $\vec{\hat{\gamma}} = (1 + \eta_0/2) \vec{\gamma}$. This replacement incorporates the modification in the Dirac equation for the modified Dirac spinors and the modification in the gauge interaction. Using
\be
\lbrace\hat{\gamma}^\mu, \hat{\gamma}^\nu\rbrace = 2 \hat{\eta}^{\mu\nu}
\ee
where $\hat{\eta}^{\mu\nu}$ is the modified Minkowski metric $\hat{\eta}^{00}=\eta^{00}=1$, $\hat{\eta}^{ii}=(1+\eta_0)\eta^{ii}=-(1+\eta_0)$ one can show that
\be
\mathrm{Tr}\left[\hat{\gamma}^{\mu_1} \hat{\gamma}^{\mu_2} \hat{\gamma}^{\mu_3} \hat{\gamma}^{\mu_4}\right] = 4 \left(\hat{\eta}^{\mu_1\mu_2}  \hat{\eta}^{\mu_3\mu_4} - \hat{\eta}^{\mu_1\mu_3}  \hat{\eta}^{\mu_2\mu_4} + \hat{\eta}^{\mu_1\mu_4} \hat{\eta}^{\mu_2\mu_3}\right).
\ee
The coefficient of $B_{\mu\nu}(k)$ in the dynamical matrix element can be read in this case from
\be
p_{1\rho} p_{\sigma} \left[\hat{\eta}^{\rho\alpha} \hat{\eta}^{\sigma\beta} -
\hat{\eta}^{\rho\sigma} \hat{\eta}^{\alpha\beta} + \hat{\eta}^{\rho\beta} \hat{\eta}^{\sigma\alpha}\right] \left(\delta_\alpha^\mu \delta_\beta^\nu -
\eta_{\alpha\beta} \eta^{\mu\nu} + \delta_\alpha^\nu \delta_\beta^\mu\right)
\label{coefftilde2}
\ee
which is the symmetrized coefficient of $p_{2\mu} p_{3\nu}$ in the product of the two traces. By contracting indices one finds
\be
2 \hat{p}^\mu \hat{p}_1^\nu + 2 \hat{p}_1^\mu \hat{p}^\nu + \left(4+3\eta_0\right) (\hat{p}_1\cdot p) \eta^{\mu\nu} - 2 (\hat{p}_1\cdot \hat{p}) \eta^{\mu\nu} - 2 (\hat{p}_1\cdot p) \hat{\eta}^{\mu\nu}\,,
\ee
to be compared with the SR coefficient
\be
p_{1\rho} p_{\sigma} \left[\eta^{\rho\alpha} \eta^{\sigma\beta} -
\eta^{\rho\sigma} \eta^{\alpha\beta} + \eta^{\rho\beta} \eta^{\sigma\alpha}\right] \left(\delta_\alpha^\mu \delta_\beta^\nu -
\eta_{\alpha\beta} \eta^{\mu\nu} + \delta_\alpha^\nu \delta_\beta^\mu\right) =
2 p^\mu p_1^\nu + 2 p_1^\mu p^\nu \,.
\label{coeffSR}
\ee
Then the dynamical matrix element will be
\be
\begin{split}
\hat{A}^{\mu\nu}(p, p_1) B_{\mu\nu}(k) &= \frac{G_F^2}{12 \pi} \left[(1-2s_W^2)^2 + (2s_W^2)^2 \right] \,\left[-4 (\hat{p}_1\cdot \hat{p}) k^2 + 4 (\hat{p}_1\cdot k) (\hat{p}\cdot k)  \right. \\ & \quad + \left. 2 \left(4+3\eta_0\right) (\hat{p}_1\cdot p) k^2 - 2 (\hat{p}_1\cdot p) (\hat{k}\cdot k)\right]\, \theta(k^2)\, \theta(k_0) \, ,
\label{tildetilde}
\end{split}
\ee
with
\begin{alignat}{2}
\hat{p} &= \left(|\vec{p}| (1+\eta_0/2), \vec{p} (1+\eta_0)\right) & \qquad
p &= \left(|\vec{p}| (1+\eta_0/2), \vec{p}\right) \nonumber \\
\hat{p}_1 &= \left(|\vec{p}_1| (1+\eta_0/2), \vec{p}_1 (1+\eta_0)\right) & \qquad
p_1 &= \left(|\vec{p}_1| (1+\eta_0/2), \vec{p}_1\right) \nonumber \\
\hat{k} &= \left((|\vec{p}|-|\vec{p}_1|) (1+\eta_0/2), (\vec{p}-\vec{p}_1) (1+\eta_0)\right) & \qquad
k &= \left((|\vec{p}|-|\vec{p}_1|) (1+\eta_0/2), (\vec{p}-\vec{p}_1)\right) \,.
\end{alignat}
At first order in an expansion in powers of the corrections to SR one has
\begin{align}
2 (\hat{p}_1\cdot \hat{p}) &= 2 |\vec{p}| |\vec{p}_1| (1-\cos\theta_1) -
2 |\vec{p}| |\vec{p}_1| \eta_0 \nonumber \\
k^2 &= - 2 |\vec{p}| |\vec{p}_1| (1-\cos\theta_1) + (|\vec{p}|-|\vec{p}_1|)^2 \eta_0 \nonumber \\
2 (\hat{p}_1\cdot k) &= 2 |\vec{p}| |\vec{p}_1| (1-\cos\theta_1) \nonumber \\
2 (\hat{p}\cdot k) &= - 2 |\vec{p}| |\vec{p}_1| (1-\cos\theta_1) \nonumber \\
2 (\hat{p}_1\cdot p) &= 2 |\vec{p}| |\vec{p}_1| (1-\cos\theta_1) \nonumber \\
(\hat{k}\cdot k) &= - 2 |\vec{p}| |\vec{p}_1| (1-\cos\theta_1)
\end{align}
and then
\begin{align}
\hat{F}_0(x_1) &= 4 x_1 (1-x_1)^2 \eta_0^2\,, \nonumber \\
\hat{F}_1(x_1) &= \left[4 x_1 (1-x_1)^2 - 8 x_1^2\right] \eta_0 \,, \nonumber \\
\hat{F}_2(x_1) &= - 8 x_1^2\,.
\end{align}
After integration on the angle $\theta_1$, we get
\be
\hat{\Gamma} = \frac {G_F^2 |\vec{p}|^5}{192 \, \pi^3} \left[(1-2s_W^2)^2 + (2s_W^2)^2 \right] \, \eta_0^3\,  \int _0^1 dx_1 \left[x_1 (1-x_1)^4 + \frac{1}{6} (1-x_1)^6\right],
\label{Gammatildetilde}
\ee
which is a third candidate for the pair production decay width, this one limited to the case of a superluminal neutrino with a momentum independent speed $v=1+\eta_0/2$.

\section{Choices of dynamical matrix element and modified dispersion relation from a field theoretical perspective}

\label{SecChoices}

From a field theoretical perspective one should start by considering a generalization of the Lagrangian of the SM containing the neutrino field. From the free part (quadratic in the neutrino field) one could read the generalization of the Dirac equation and the corresponding modified dispersion relation (i.e. the function $\epsilon(|\vec{p}|)$). By also considering the interaction term (product of two neutrino fields and the $Z$-boson field) one could derive the expression for the dynamical matrix element ($\sqamp$).

Since one is considering a generalization of a relativistic gauge theory the most natural way to implement a modified dispersion relation in an extended Lagrangian is to add a new term in the free fermion action that fixes the modification in the dispersion relation and to consider the same dynamical gauge principle (replacing partial by covariant derivatives) that fixes the gauge interaction in the relativistic limit. The fourth example for the dynamical matrix element considered in the previous section is just the simplest choice along these lines with just one spatial derivative in the free part. One could consider generalizations with higher spatial derivative terms that would lead to new interaction terms and new contributions to the dynamical matrix element. A generalization along these lines is restricted to the effective field theory framework (derivative expansion) so that the choice of a modified dispersion relation is very limited and the study of implications of (or constraints on) such a generalization is easier than in other cases. To be precise, the gauge invariance that has been implemented in the Lorentz violating generalization of the SM Lagrangian in Eq.~(\ref{Leta0}) is a $U(1)$ symmetry instead of the complete $SU_L(2)\times U_Y(1)$ gauge invariance of the Lorentz invariant contribution. Should we had considered an $SU_L(2)\times U_Y(1)$ gauge invariant extended Lagrangian, Lorentz violations on different particles would be strongly restricted. In particular since the neutrino field and the left-handed charged lepton field are in a gauge doublet then Lorentz violations in the neutrino sector automatically would have an analogue in the charged lepton sector. The absence of observations of such Lorentz violating effects for electrons implies strong restrictions on posible effects due to Lorentz violations in neutrino physics including the possible energy loss of superluminal neutrinos that we are discussing in detail in this work.

An alternative way to implement a modified dispersion relation is to consider an extension of the relativistic Lagrangian independent of the gauge fields constructed from gauge invariant products of matter fields.\footnote{The consistency of this field theoretical framework in the presence of radiative corrections can be questioned~\cite{Giudice:2011mm} and deserves further study.} This possibility is restricted by the matter field content of the theory. In the case of the SM Lagrangian the fact that the (conjugate of the) doublet scalar field $\tilde{\Phi}$ and the left-handed doublet lepton field $L$
\be
\tilde{\Phi} =  \left(\begin{array}{c}\phi^{0*} \\
  \phi^{-}\end{array}\right)   {\hskip 2cm} L = \left(\begin{array}{c}\nu_{L} \\
  l_{L}\end{array}\right)
\ee
have the same gauge transformations allow to consider an extended Lagrangian quadratic in the gauge invariant product of these two field doublets
\be
\mathcal{L}_{LIV}(\tilde{\Phi}^\dagger L) = \frac{1}{M^2} \left(\bar{L} \tilde{\Phi}\right) i \vec{\gamma} \cdot \vec{\nabla} \, \epsilon(|i\vec{\nabla}|) \left(\tilde{\Phi}^\dagger L\right) \, .
\ee
In the approximation where one neglects the fluctuations in the scalar field this extended Lagrangian reduces to a quadratic Lagrangian in the neutrino left-handed field. This opens the possibility to consider Lorentz violating effects in the neutrino sector with no analogue for other particles so that restrictions from the absence of observations of such effects in other systems do not translate directly into restrictions on possible effects of Lorentz violations in the neutrino sector. This provides us with an example where neutrino physics is a special window to explore departures from SR. As far as one does not require a perturbative treatment of the Lorentz violating interactions in the lepton-scalar sector there is not any restriction on the quadratic extended Lagrangian so that one can consider arbitrary choices for the modified neutrino dispersion relation (arbitrary choice for $\epsilon(|\vec{p}|$)) going beyond a derivative expansion (momentum power expansion). The second example (\ref{A2}) for the matrix element of pair production by superluminal neutrinos can be seen as a result within this framework.

A third alternative to the generalization of a relativistic gauge theory would correspond to assume that the local gauge symmetry is a property of the relativistic limit. In this case one could consider an extended Lagrangian with no restrictions from local gauge invariance. The second example (\ref{A2}) could also be seen as a tree level approximation within this framework which can not be distinguished from the second alternative unless one goes beyond this approximation. An argument in favor of the realization of the second example for the dynamical matrix element within the previous second alternative for a generalized relativistic gauge theory is that it is not clear how the introduction of Lorentz violating terms in a Lagrangian provides a way to escape to the inconsistencies of a gauge non-invariant relativistic field theory with vector fields.

The first example (\ref{A1}) for the dynamical matrix element can not be derived from a field theory perturbative calculation. It therefore requires to consider a generalization of the relativistic gauge theory that goes beyond the field theory framework and then there is no reason to consider restrictions on the choice of modified dispersion relation.

For the second example (\ref{A2}) one can consider a momentum power expansion for the modified dispersion relation ($\epsilon(|\vec{p}|$) if one assumes the validity of the effective field theory framework for the study of Lorentz violating effects or a more general momentum dependence if one assumes that one has to go beyond the effective field theory expansion when one goes beyond the special relativistic limit.

The results for the third example, although can not be derived from a field theory calculation, are equivalent to those of the second example at leading order in the deviations from SR and then do not require any independent discussion until one goes beyond the leading order effects.

Finally in the fourth example for the dynamical matrix element there is a definite modified dispersion relation corresponding to a momentum independent speed (in the massless limit). We could go beyond this case by including gauge invariant higher derivative terms in the Lagrangian (\ref{Leta0}).

\section{Energy loss of superluminal neutrinos}
\label{SecEnergy}
In this section we evaluate the width for pair production, the rate of energy loss, and the energy of a superluminal neutrino after propagation over a given distance for different choices for the dynamical matrix element and the modified dispersion relation. These results are the starting point of an analysis of the observable consequences of having superluminal neutrinos with the uncertainties due to the lack of knowledge of the details of the possible origin of the Lorentz invariance violation in neutrino physics.

\subsection{Decay width}
We have three candidates (\ref{Gamma}), (\ref{Gammatilde}), (\ref{Gammatildetilde}) for the decay width as a functional of the modified dispersion relation corresponding to different choices for the dynamical matrix element of the pair production process.

If we consider the simplest choice for the modified dispersion relation $\epsilon(|\vec{p}|)= \eta_0/2$ (momentum independent speed) with $\eta_0 > 0$ (for negative values of $\eta_0$ the decay width is zero) then the threshold is
$|\vec{p}_{th}|=\sqrt{4\, m_e^2/\eta _0}$ and the decay widths are:
\begin{align}
\Gamma_0 &= \frac{G_F^2 |\vec{p}|^5}{192 \, \pi^3} \left[(1-2s_W^2)^2 + (2s_W^2)^2 \right] \, \xi_0  \left(\frac{\eta_0}{2}\right)^3 \label{Gamma_0} \\
\tilde{\Gamma}_0 &=  \frac{G_F^2 |\vec{p}|^5}{192 \, \pi^3} \left[(1-2s_W^2)^2 + (2s_W^2)^2 \right] \, \tilde{\xi}_0 \left(\frac{\eta_0}{2}\right)^3 \label{Gammatilde_0} \\
\hat{\Gamma}_0 &= \frac{G_F^2 |\vec{p}|^5}{192 \, \pi^3} \left[(1-2s_W^2)^2 + (2s_W^2)^2 \right] \, \hat{\xi}_0 \left(\frac{\eta_0}{2}\right)^3 \label{Gammatildetilde_0}
\end{align}
with
\be
\xi_0 = \frac{8}{7}  {\hskip 2cm}
\tilde{\xi}_0 = \frac{34}{105}  {\hskip 2cm}
\hat{\xi}_0 = \frac{16}{35}\,.
\label{gamma0}
\ee
The result for the decay width (\ref{Gamma_0}) corresponding to the first choice for the dynamical matrix element (SR matrix element with the replacement of the SR energy by their modified expression in terms of the momentum) reproduces the result used in Ref.~\cite{Cohen:2011hx} to argue against the consistency of the recent result of OPERA~\cite{Adam:2011zb} for a superluminal neutrino velocity.
The other two results in Eq.~(\ref{gamma0}) also reproduce the two cases studied in Ref.~\cite{Bez}, which in fact can be seen to correspond to the second and fourth matrix elements of the previous section for a momentum independent velocity. Our more direct computation will however allow us to consider other dispersion relations beyond the constant speed case.

We can see from (\ref{gamma0}) that a change in the choice of the dynamical matrix element produces an additional overall factor of $(17/60)$ in the decay width if one takes the dynamical matrix element corresponding to a perturbative field theory calculation with the simplest free action implementing a modified Dirac equation and the SR interaction (second example of the previous section) or an overall factor of $(2/5)$ when one uses an interaction fixed by the dynamical gauge invariance principle (fourth example).

The next choice we consider for the modification in the dispersion relation is
$\epsilon(|\vec{p}|)= |\vec{p}|^n/\Lambda^n$ (with $n$ and $\Lambda$ positive numbers) which corresponds to a
Lorentz violating free term in the Lagrangian with $n$ spatial derivatives when $n$ is a natural number. The energy scale $\Lambda$ is the UV scale that fixes the domain of validity of the effective field theory energy power expansion. In this case, the threshold is given by $|\vec{p}_{th}|^{n+2}= 2\, m_e^2 \, \Lambda ^n$ and the widths are:
\begin{align}
\Gamma_n &= \frac {G_F^2 |\vec{p}|^{5}}{192 \, \pi^3} \left[(1-2s_W^2)^2 + (2s_W^2)^2 \right] \, \left(\frac{|\vec{p}|}{\Lambda}\right)^{3n} \, \xi_n   \label{Gamma_n} \\
\tilde{\Gamma}_n &= \frac {G_F^2 |\vec{p}|^{5}}{192 \, \pi^3} \left[(1-2s_W^2)^2 + (2s_W^2)^2 \right] \, \left(\frac{|\vec{p}|}{\Lambda}\right)^{3n} \, \tilde{\xi}_n \label{Gammatilde_n}
\end{align}
with
\begin{align}
\xi_n &= 2 - \frac{12 (n+6)}{(n+3)(n+4)(n+5)} + \frac{12 (n+1)}{(n+2)(2n+3)(2n+5)} - \frac{4(3n+2)}{(3n+4)(3n+5)(3n+7)} \nonumber \\
\tilde{\xi}_n &= 1 - \frac{12}{(n+2)(n+5)} + \frac{6}{(n+3)(2n+3)} - \frac{4}{(3n+4)(3n+7)}\,.
\end{align}
The main difference between (\ref{Gamma_n})-(\ref{Gammatilde_n}) and the results (\ref{Gamma_0})-(\ref{Gammatilde_0}) for the simplest choice for the dispersion relation is the power exponent in the momentum dependence of the decay width. This makes the effect of the production of pairs in the propagation of neutrinos to increase much faster when the energy increases.

As a third choice for the modification of the dispersion relation at high energies ($|\vec{p}|>p_0$), we use $\epsilon(|\vec{p}|) = \lambda^\alpha/|\vec{p}|^\alpha$ with\footnote{If $\alpha>1$ then one has subluminal neutrinos and no pair production.} $0<\alpha\leq1$ and $\lambda$ an additional energy scale required by dimensional arguments. This corresponds to a nonlocal free field theory action trying to illustrate a Lorentz violation in the neutrino sector that goes beyond the effective field theory framework. In this case, the threshold is given by
$|\vec{p}_{th}|^{2-\alpha}= 2\, m_e^2 /\lambda ^{\alpha}$ and the widths (for $|\vec{p}|\gg |\vec{p}_0|$) are:
\begin{align}
\Gamma_{-\alpha} &=  \frac {G_F^2 |\vec{p}|^{5}}{192 \, \pi^3} \left[(1-2s_W^2)^2 + (2s_W^2)^2 \right] \, \left(\frac{\lambda}{|\vec{p}|}\right)^{3\alpha} \, \xi_{-\alpha}   \label{Gamma-alpha} \\
\tilde{\Gamma}_{-\alpha} &=  \frac {G_F^2 |\vec{p}|^{5}}{192 \, \pi^3} \left[(1-2s_W^2)^2 + (2s_W^2)^2 \right] \, \left(\frac{\lambda}{|\vec{p}|}\right)^{3\alpha} \, \tilde{\xi}_{-\alpha}
\label{tildeGamma-alpha}
\end{align}
with
\begin{align}
\xi_{-\alpha} &= 2 - \frac{12 (6-\alpha)}{(3-\alpha)(4-\alpha)(5-\alpha)} + \frac{12 (1-\alpha)}{(2-\alpha)(3-2\alpha)(5-2\alpha)} - \frac{4(2-3\alpha)}{(4-3\alpha)(5-3\alpha)(7-3\alpha)} \nonumber \\
\tilde{\xi}_{-\alpha} &= 1 - \frac{12}{(2-\alpha)(5-\alpha)} + \frac{6}{(3-\alpha)(3-2\alpha)} - \frac{4}{(4-3\alpha)(7-3\alpha)}\,.
\end{align}
In this case the increase of the decay width of pair production with the energy is slower than in the case of a momentum independent velocity of propagation. In addition, note that in the expression of the decay rate there are energy-independent, dimensionless factors (i.e., pure numbers) that depend on the particular form of the dispersion relation and which can differ by one order of magnitude. For example,
\begin{equation}
\dfrac{\tilde \xi_{2}}{\tilde \xi_{-3/4}}\simeq 18.8 \, .
\end{equation}

As a last illustrative example, we shall consider a case in which the dispersion relation is non-analytic and it is given by $\epsilon(|\vec{p}|) = \, e^{-\Lambda/|\vec{p}|}$. In this case, that also goes beyond EFT, the decay width is extremely sensitive to changes in the momentum. When $|\vec{p}|\ll\Lambda$, the dominant contribution to the decay width for the second matrix element is:
\be
\tilde{\Gamma} = \frac{G_F^2 |\vec{p}|^5}{192 \pi^3} \, \left[(1-2s_W^2)^2 + (2s_W^2)^2 \right] \,  e^{-3\Lambda/|\vec p |} \, ;
\ee
While this expression is small in the region in which is valid, it illustrates the possibility of having a remarkably strong dependence on momentum.

\subsection{Rate of energy loss}
An approximation to the effect of the production of $e^+e^-$ pairs on the propagation of neutrinos can be obtained from the rate of energy loss
\be
\frac{d|\vec{p}|}{dt} = - |\vec{p}| \, \int_0^1 dx_1 (1-x_1) \frac{d\Gamma}{dx_1} \,.
\ee
In the case of $\epsilon(|\vec{p}|) = \eta_0/2$ we have
\begin{align}
\frac{d|\vec{p}|}{dt} &= - \frac{G_F^2 |\vec{p}|^6}{192 \, \pi^3} \left[(1-2s_W^2)^2 + (2s_W^2)^2 \right] \, {\xi'_0} \, \left(\frac{\eta_0}{2}\right)^3 \\
\frac{d|\vec{p}|}{dt} &= - \frac{G_F^2 |\vec{p}|^6}{192 \, \pi^3} \left[(1-2s_W^2)^2 + (2s_W^2)^2 \right] \, {\tilde{\xi'}_0} \left(\frac{\eta_0}{2}\right)^3 \\
\frac{d|\vec{p}|}{dt} &= - \frac{G_F^2 |\vec{p}|^6}{192 \, \pi^3} \left[(1-2s_W^2)^2 + (2s_W^2)^2 \right] \, {\hat{\xi'}_0} \left(\frac{\eta_0}{2}\right)^3
\end{align}
with
\be
{\xi'_0} = \frac{25}{28}  {\hskip 2cm}  {\tilde{\xi'}_0} = \frac{11}{42}  {\hskip 2cm} {\hat{\xi'}_0} = \frac{5}{14}
\label{xiprimas}
\ee
for the decay width results in (\ref{Gamma}), (\ref{Gammatilde}), and (\ref{Gammatildetilde}), respectively.

If one considers $\epsilon(|\vec{p}|)= |\vec{p}|^n/\Lambda^n$ then one has
\begin{align}
\frac{d|\vec{p}|}{dt} &= - \frac{G_F^2 |\vec{p}|^{6}}{192 \, \pi^3} \left[(1-2s_W^2)^2 + (2s_W^2)^2 \right] \, \left(\frac{|\vec{p}|}{\Lambda}\right)^{3n} \, {\xi'_n} \\
\frac{d|\vec{p}|}{dt} &= - \frac{G_F^2 |\vec{p}|^{6}}{192 \, \pi^3} \left[(1-2s_W^2)^2 + (2s_W^2)^2 \right] \, \left(\frac{|\vec{p}|}{\Lambda}\right)^{3n} \, {\tilde{\xi'}_n}
\label{dpdtn}
\end{align}
with
\be
\begin{split}
{\xi'_n} &= \frac{13}{10} - \frac{12 (2n+15)}{(n+3)(n+4)(n+5)(n+6)} + \frac{12(4n+3)}{(2n+3)(2n+4)(2n+5)(2n+6)} \\ & \quad - \frac{2(5n+4)}{(3n+4)(3n+5)(3n+6)} + \frac{10}{3(3n+7)(3n+8)}  \\
\end{split}
\ee
\be
\begin{split}
{\tilde{\xi'}_n} &= \frac{3}{5} - \frac{24(n+4)}{(n+2)(n+3)(n+5)(n+6)} +
\frac{24(2n+5)}{(2n+3)(2n+4)(2n+6)(2n+7)} \\ & \quad - \frac{8(3n+6)}{(3n+4)(3n+5)(3n+7)(3n+8)}\,.
\end{split}
\ee

With the third choice of modified dispersion relation, $\epsilon(|\vec{p}|) = \lambda^\alpha/|\vec{p}|^\alpha$, the rate of energy loss is given by
\begin{align}
\frac{d|\vec{p}|}{dt} &= - \frac{G_F^2 |\vec{p}|^{6}}{192 \, \pi^3} \left[(1-2s_W^2)^2 + (2s_W^2)^2 \right] \, \left(\frac{\lambda}{|\vec{p}|}\right)^{3\alpha} \, {\xi'_{-\alpha}}  \nonumber \\
\frac{d|\vec{p}|}{dt} &= - \frac{G_F^2 |\vec{p}|^{6}}{192 \, \pi^3} \left[(1-2s_W^2)^2 + (2s_W^2)^2 \right] \, \left(\frac{\lambda}{|\vec{p}|}\right)^{3\alpha} \, {\tilde{\xi'}_{-\alpha}}
\label{dpdt-alpha}
\end{align}
with
\be
\begin{split}
{\xi'_{-\alpha}} & =  \frac{13}{10} - \frac{12 (15-2\alpha)}{(3-\alpha)(4-\alpha)(5-\alpha)(6-\alpha)} + \frac{12(3-4\alpha)}{(3-2\alpha)(4-2\alpha)(5-2\alpha)(6-2\alpha)} \\ & \quad - \frac{2(4-5\alpha)}{(4-3\alpha)(5-3\alpha)(6-3\alpha)} + \frac{10}{3(7-3\alpha)(8-3\alpha)}
\end{split}
\ee
\be
\begin{split}
{\tilde{\xi'}_{-\alpha}} &= \frac{3}{5} - \frac{24(4-\alpha)}{(2-\alpha)(3-\alpha)(5-\alpha)(6-\alpha)} +
 \frac{24(5-2\alpha)}{(3-2\alpha)(4-2\alpha)(6-2\alpha)(7-2\alpha)} \\ & \quad - \frac{8(6-3\alpha)}{(4-3\alpha)(5-3\alpha)(7-3\alpha)(8-3\alpha)}\,.
\end{split}
\ee

\subsection{Final energy after propagation}
The previous results of the rate of energy loss can be used to get an estimate of the final energy $E_f$ of a neutrino of energy $E_i$ after propagating over a distance $L$.

One has
\be
\frac{1}{E_f^5} - \frac{1}{E_i^5} = \frac{1}{E_0^5} {\hskip 2cm}
\frac{1}{E_f^5} - \frac{1}{E_i^5} = \frac{1}{\tilde{E}_0^5}  {\hskip 2cm}
\frac{1}{E_f^5} - \frac{1}{E_i^5} = \frac{1}{\hat{E}_0^5}
\ee
with
\be
\tilde{E}_0 = \left[\frac{G_F^2 L}{192 \, \pi^3} \left[(1-2s_W^2)^2 + (2s_W^2)^2 \right] \, 5 \, \left(\frac{\eta_0}{2}\right)^3  \, {\tilde{\xi}'_0}\right]^{-1/5}
\label{E0}
\ee
in the case of a momentum independent velocity ($\epsilon(|\vec{p}|)=\eta_0/2$). The result for $E_0$ ($\hat{E}_0$) is obtained from (\ref{E0}) by the replacement of ${\tilde{\xi'}_0}$ by ${\xi'_0}$ (${\hat{\xi'}_0}$).

For a modification of the dispersion relation with $\epsilon(|\vec{p}|)=|\vec{p}|^n/\Lambda^n$ one has
\be
\frac{1}{E_f^{5+3n}} - \frac{1}{E_i^{5+3n}} = \frac{1}{E_n^{5+3n}} {\hskip 2cm}
\frac{1}{E_f^{5+3n}} - \frac{1}{E_i^{5+3n}} = \frac{1}{\tilde{E}_n^{5+3n}}
\ee
with
\be
{\tilde E}_n =  \left[\frac{G_F^2 L}{192 \, \pi^3} \left[(1-2s_W^2)^2 + (2s_W^2)^2 \right] \, (5+3n) \,  \frac{1}{\Lambda^{3n}} \, {{\tilde \xi}'_n}\right]^{-1/(5+3n)}
\label{En}
\ee
and a similar result for $E_n$ replacing ${\tilde{\xi'}_n}$ by ${\xi'_n}$.

Finally, in the case $\epsilon(|\vec{p}|) = \lambda^\alpha/|\vec{p}|^\alpha$ one has
\be
\frac{1}{E_f^{5-3\alpha}} - \frac{1}{E_i^{5-3\alpha}} = \frac{1}{E_{-\alpha}^{5-3\alpha}} {\hskip 2cm}
\frac{1}{E_f^{5-3\alpha}} - \frac{1}{E_i^{5-3\alpha}} = \frac{1}{\tilde{E}_{-\alpha}^{5-3\alpha}}
\ee
with
\be
\tilde{E}_{-\alpha} = \left[\frac{G_F^2 L}{192 \, \pi^3} \left[(1-2s_W^2)^2 + (2s_W^2)^2 \right] \, (5-3\alpha) \, \lambda^{3\alpha} \, {\tilde{\xi}'_{-\alpha}}\right]^{-1/(5-3\alpha)}
\label{E-alpha}
\ee
and $E_{-\alpha}$ with a factor ${\xi'_{-\alpha}}$ instead of ${\tilde{\xi'}_{-\alpha}}$.

\subsection{Some numerical estimates}

We can take the inverse of the decay width of pair production as an estimate of the distance that neutrinos should propagate to have an appreciable loss of energy. From (\ref{Gammatilde_0}) we have
\be
\tilde{\Gamma}_0^{-1} = \frac{4.3 \, \times \, 10^{-4}}{\eta_0^3} \, \left(\frac{\text{GeV}}{|\vec{p}|}\right)^5 \, \text{km}
\label{Gamma0-km}
\ee
and similar results for $\Gamma_0^{-1}$ ($\hat{\Gamma}_0^{-1}$) with an additional factor $17/60$ ($17/24$) in the inverse of the decay width.
In the case of a Lorentz violating free term with $n$ spatial derivatives one has
\be
\tilde{\Gamma}_n^{-1} = \frac{1.7 \, \times \, 10^{(57n-5)}}{\tilde{\xi}_n} \, \left(\frac{\Lambda}{10^{19} \, \text{GeV}}\right)^{3n} \, \left(\frac{\text{GeV}}{|\vec{p}|}\right)^{(5+3n)} \, \text{km}
\label{Gamman-km}
\ee
and in the case of a Lorentz violating correction which decreases at large momenta (\ref{tildeGamma-alpha}),
\be
\tilde{\Gamma}_{-\alpha}^{-1} = \frac{1.7 \, \times \, 10^{(27\alpha-5)}}{\tilde{\xi}_{-\alpha}} \,
\left(\frac{\text{eV}}{\lambda}\right)^{3\alpha} \, \left(\frac{\text{GeV}}{|\vec{p}|}\right)^{(5-3\alpha)} \, \text{km} \,.
\label{Gamma-alpha-km}
\ee
These expressions give us an idea of the sensitivity to an energy-loss due to pair production in the propagation of neutrinos from observations of solar, reactor, accelerator or atmospheric neutrinos.

The results for the energy scales (\ref{E0}), (\ref{En}) and (\ref{E-alpha}),
\begin{align}
\tilde{E}_0 &= \frac{1.6 \, \times \, 10^{-1}}{\eta_0^{3/5}} \, \left(\frac{\text{km}}{L}\right)^{1/5} \, \text{GeV}
\label{E0-km} \\
\tilde{E}_n &= \left(\frac{1.7 \, \times \, 10^{(57n-5)}}{(5+3n){\tilde{\xi}'_n}}\right)^{1/(5+3n)} \, \left(\frac{\Lambda}{10^{19} \, \text{GeV}}\right)^{3n/(5+3n)} \, \left(\frac{\text{km}}{L}\right)^{1/(5+3n)} \, \text{GeV}
\label{En-km} \\
\tilde{E}_{-\alpha}& = \left(\frac{1.7 \, \times \, 10^{(27\alpha-5)}}{(5-3\alpha){\tilde{\xi}'_{-\alpha}}}\right)^{1/(5-3\alpha)} \, \left(\frac{\text{eV}}{\lambda}\right)^{3\alpha/(5-3\alpha)} \, \left(\frac{\text{km}}{L}\right)^{1/(5-3\alpha)} \, \text{GeV}
\label{E-alpha-km}
\end{align}
which allow us to determine the final energy after a propagation over a distance $L$, give us also another estimate of the possible constraints that one can get on (or hints of) Lorentz violating corrections to neutrino physics from the observed high energy neutrino spectrum.

The results in (\ref{E0-km})-(\ref{E-alpha-km}) correspond to our second example (\ref{Gammatilde}) for the dynamical matrix element. For the other examples of matrix elements one has
\be
E_0 = \tilde{E}_0 \, \left(\frac{{\tilde{\xi'}_0}}{\xi'_0}\right)^{1/5}
{\hskip 1cm} \hat{E}_0 = \tilde{E}_0 \, \left(\frac{\tilde{\xi}'_0}{{\hat{\xi'}_0}}\right)^{1/5}
{\hskip 1cm} E_n = \tilde{E}_n \, \left(\frac{{\tilde{\xi'}_n}}{\xi'_n}\right)^{1/(5+3n)}
{\hskip 1cm} E_{-\alpha} = \tilde{E}_{-\alpha} \, \left(\frac{{\tilde{\xi'}_{-\alpha}}}{\xi'_{-\alpha}}\right)^{1/(5-3\alpha)}.
\ee

\begin{figure}
\centerline{\includegraphics[scale=0.8]{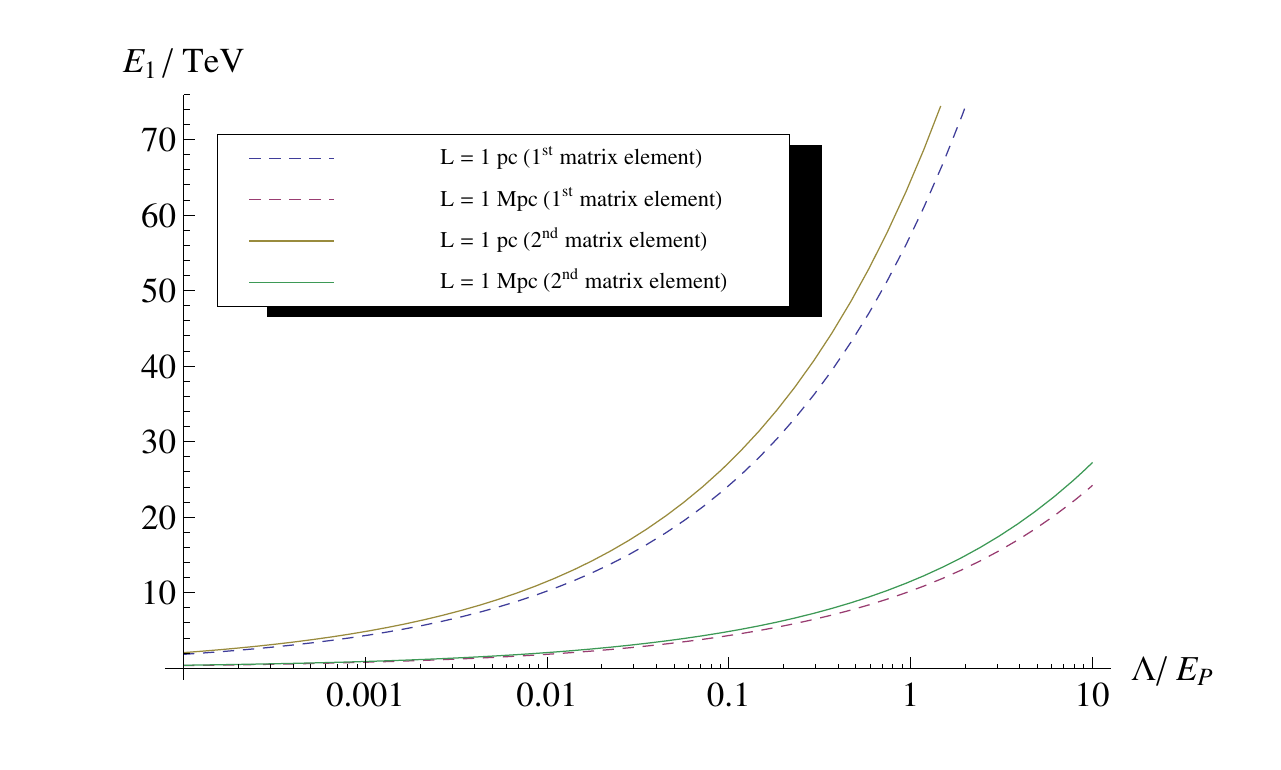} \hspace{-1cm} \includegraphics[scale=0.8]{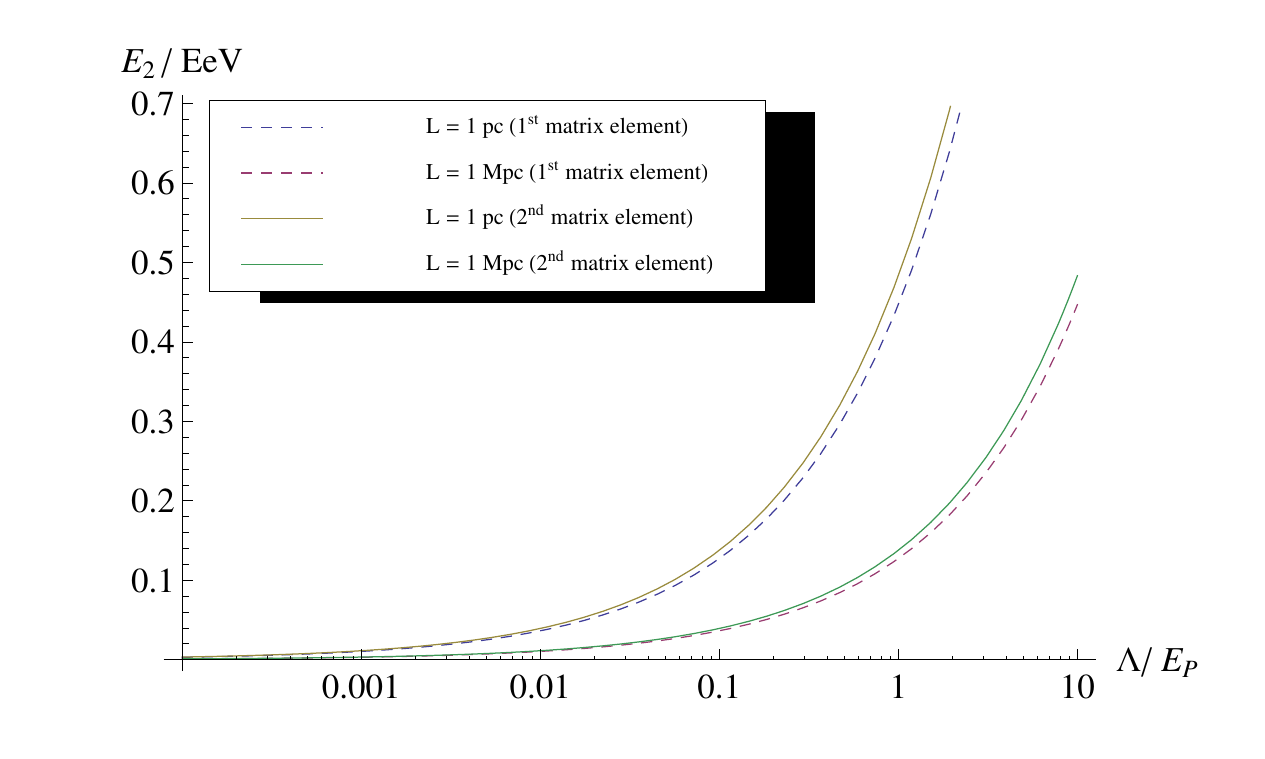}}
\caption{Value of the terminal energy $E_n$ (approximate final energy of a neutrino with initial energy higher than $E_n$) for the cases $n=1$ (left) and $n=2$ (right) as a function of the high-energy scale $\Lambda$ written in terms of the Planck energy $E_P$. The curves show the differences with respect to the matrix element used in the calculation and with different distances of propagation, typical of galactic or extragalactic neutrinos.}
\label{fig:galactic}
\end{figure}

For galactic and extragalactic neutrinos, it is convenient to change units and rewrite, e.g., Eq.~(\ref{Gamma0-km}) as
\be
\tilde{\Gamma}_0^{-1} = \frac{1.4 \, \times \, 10^{-32}}{\eta_0^3} \, \left(\frac{\text{TeV}}{|\vec{p}|}\right)^5 \, \text{pc}
\label{Gamma0-pc}
\ee
and
\be
\tilde{\Gamma}_0^{-1} = \frac{1.4 \, \times \, 10^{-68}}{\eta_0^3} \, \left(\frac{10^9 \, \text{GeV}}{|\vec{p}|}\right)^5 \, \text{Mpc} \,.
\label{Gamma0-Mpc}
\ee
As an illustrative example, Fig.~\ref{fig:galactic} gives the value of $\tilde{E}_n$ for the two most studied cases in the literature of quantum gravity phenomenology, $n=1$ and $n=2$, in the case of galactic (propagation distance of the order of pc) and extragalactic (propagation distance of the order of Mpc) neutrinos. We also show in that figure the differences in the use of the first or second matrix elements in the calculation.

\section{Times of flight of superluminal neutrinos}
\label{SecTimes}
A modification of the neutrino dispersion relation produces an energy loss in the propagation of neutrinos related with the time of flight. Then one can try to look for neutrino observations where one can simultaneously determine the energy loss and time of flight in order to test the consistency of the observations with a given modification of the dispersion relation. The determination of the energy loss is limited by the uncertainties of our knowledge of the sources of very high energy neutrinos. As for the time of flight determination one has together with the uncertainties in the knowledge of the time of emission also the uncertainties due to the limited precision of time measurements.

In the case of a momentum independent superluminal velocity it is straightforward to calculate the time of flight of a neutrino propagating over a distance $L$
\be
\Delta t_0 =  3.3 \times 10^3 \, \left(\frac{L}{\text{km}}\right) \, \left(1-\frac{\eta_0}{2}\right) \, \text{ns}
\label{t0-ns}
\ee
or
\be
\Delta t_0 = 1.3 \times 10^{14} \, \left(\frac{L}{\text{Mpc}}\right) \, \left(1-\frac{\eta_0}{2}\right) \, \text{s} \,.
\label{t0-s}
\ee
A deviation, with respect to the SR expectation, in the time of flight of neutrinos of the order of $10 \, \text{ns}$ in the propagation over a distance of the order of $10^3 \, \text{km}$ would correspond to $\eta_0 \sim 10^{-5}$. From (\ref{Gamma0-km}) one can see that an inverse decay length of $10^3 \, \text{km}$ for such a value of $\eta_0$ corresponds to a neutrino with $|\vec{p}| \sim 40 \,\text{GeV}$. This provides us with a quantitative comparison of the sensitivity of neutrino times of flight and spectrum observations to a departure from SR kinematics with a momentum independent speed for the neutrinos. Times of flight of extragalactic neutrinos are much more sensitive to deviations from SR with a deviation of the order of a few $\text{s}$ for a neutrino propagating over a $\text{Mpc}$ corresponding to much smaller values of $\eta_0$ ($\eta_0 \sim 10^{-13}$). An inverse decay width of a $\text{Mpc}$ with this value of $\eta_0$ corresponds to a neutrino with $|\vec{p}| \sim \text{TeV}$. An extragalactic neutrino flux extending above these energies excludes then such a deviation from SR.

In the case of more general modifications of the dispersion relation the time of flight of neutrinos will be affected by the energy loss due to the production of $e^+e^-$ pairs. The momentum and then the velocity of the neutrino is changing in the propagation according to (\ref{dpdtn}) or (\ref{dpdt-alpha}). It is very easy to calculate in these two cases the neutrino time of flight over a distance $L$. In the case of $\epsilon(|\vec{p}|)=|\vec{p}|^n/\Lambda^n$ one has
\be
\Delta t_n = L - L \, (n+1) \left(\frac{E_f}{\Lambda}\right)^n \, \tau_n
\ee
with
\be
\tau_n = \frac{(5+3n)}{(5+2n)} \, \left(\frac{E_n}{E_f}\right)^{(5+3n)} \, \left(1 - \left[1-\left(\frac{E_f}{E_n}\right)^{(5+3n)}\right]^{(5+2n)/(5+3n)} \right)\,.
\ee
In the case of $\epsilon(|\vec{p}|)=\lambda^\alpha/|\vec{p}|^\alpha$ one has
\be
\Delta t_{-\alpha} = L - L \, (1-\alpha) \left(\frac{\lambda}{E_f}\right)^\alpha \,
\tau_{-\alpha}
\ee
with
\be
\tau_{-\alpha} = \frac{(5-3\alpha)}{(5-2\alpha)} \, \left(\frac{E_{-\alpha}}{E_f}\right)^{(5-3\alpha)} \, \left(1 - \left[1-\left(\frac{E_f}{E_{-\alpha}}\right)^{(5-3\alpha)}\right]^{(5-2\alpha)/(5-3\alpha)} \right)\,.
\ee
In the case of $E_f^5 \cdot L \ll 192 \pi ^3/G_F^2$ one has $E_f \ll E_n$ ($E_f \ll E_{-\alpha}$) and $\tau_n\approx 1$ ($\tau_{-\alpha}\approx 1$), i.e., the time of flight corresponding to a uniform motion with a momentum dependent speed.

\section{Consistency of the superluminal interpretation of OPERA observations}
\label{SecConsistency}

Recently there has been a claim of an observation of neutrinos propagating with superluminal velocities in the CNGS beam from CERN to Gran Sasso~\cite{Adam:2011zb} (OPERA experiment). It is then natural to try to accommodate these observations within the present discussion of the effect of Lorentz violations on the propagation of neutrinos.

Almost all the theoretical discussions related to OPERA consider a dispersion relation with
$\epsilon(|\vec{p}|) =\eta_0/2$. This is due to the absence of a change in the measured times of flight in all the range of energies detected going from $10$ GeV up to $100$ GeV. These results, if confirmed with more statistics, put strong constraints on any momentum dependence of the velocity of propagation of neutrinos at least in the range of energies covered by the OPERA time of flight measurements.
But in fact the OPERA value for $\eta_0=4.7\times 10^{-5}$~\cite{Adam:2011zb} is in obvious conflict\footnote{Assuming electron antineutrinos propagate with the same speed as the muon neutrinos detected by OPERA.} with time of flight limits from SN1987A~\cite{Hirata:1987hu,Bionta:1987qt,Longo:1987ub,Stodolsky:1987vd} as one can see from Eq.~(\ref{t0-s}). This conflict requires to consider a deviation from a momentum independent choice for $\epsilon$ at energies below those explored by OPERA including the range of energies of SN neutrinos (few MeV). There is also a conflict of this large value of $\eta_0$ with high energy neutrino observations owing to its implications on the propagation of high energy atmospheric neutrinos. This requires a strong suppression of the production of $e^+e^-$ pairs at energies well above those explored by OPERA with an appropriate choice of the momentum dependence of $\epsilon$.
This is also needed to avoid incompatibilities with neutrino production owing to the kinematic corrections in the pion decay induced by such large value of $\eta_0$~\cite{Cowsik:2011wv,Bi:2011nd,GonzalezMestres:2011jc}.

The simplest way to try to escape to these contradictions is to use a modified dispersion relation combining a function $\epsilon(|\vec{p}|)$ which increases when going from SN momenta to OPERA momenta, stays almost constant over the energy range of OPERA observations to be consistent with the very mild momentum dependence of the time of flight of neutrinos and then decreases if we go beyond OPERA momenta in order to escape to incompatibilities with the observed high energy neutrino spectra. However, there is still another conflict owing to the effect of the production of $e^+e^-$ pairs in the propagation of neutrinos from CERN to Gran Sasso. According to the previous section, one can see that, for the value of $\eta_0$ required to reproduce OPERA results on time of flights, there is a very drastic correction on the neutrino energies in conflict with observations.

For a momentum independent velocity of propagation in the OPERA energy range, taking Eq.~(\ref{E0-km}) with $\eta_0=4.7\times 10^{-5}$ and $L=731$\,km as the distance traveled by neutrinos in the experiment, we get $\tilde{E}_0=17.0$\,GeV in the case of the second matrix element. In the case of the first matrix element, which is the one considered by Cohen and Glashow in Ref.~\cite{Cohen:2011hx}, we obtain $E_0=13.3$\,GeV (in fact they get a slightly different result, $E_0=12.7$\,GeV, because they use $\eta_0=5.0\times 10^{-5}$, a value which was updated by the OPERA collaboration in November 2011 to $\eta_0=4.7\times 10^{-5}$, which is the one used in our calculation). For the third matrix element, $\hat{E}_0=16.0$\,GeV. This means that together with the rather adhoc choice of the Lorentz violating corrections in the free Lagrangian which would produce the peculiar behavior of the $\epsilon(|\vec{p}|)$ function as described in the previous paragraph, one has to assume a suppression in the dynamical matrix element relevant for the calculation of the energy loss of neutrinos propagating from CERN to Gran Sasso.
From Eq.~(\ref{E0}), we get
\be
E_0 = 0.123 \, \left(\frac{L}{\text{km}}\right)^{-1/5} \, \eta_0^{-3/5} \, {\xi'_0}^{-1/5}\, \text{GeV}\,.
\label{bounds}
\ee
Let us consider this relation for a generic matrix element. Since OPERA observes the arrival of neutrinos of energies higher than, let us take, $E_0\geq 60$\,GeV, Eq.~(\ref{bounds}) gives an upper bound for $\xi'_0$:
\be
\xi'_0 \leq 4.8 \times 10^{-4}.
\ee
This means that in order to make the OPERA time-of-flight measurement compatible with the non-observation of pair production, one needs a suppression in the dynamical matrix element so that $\xi'\sim \mathcal{O}(10^{-4})$ instead of the $\mathcal{O}(10^{-1})$ values of Eq.~(\ref{xiprimas}) obtained in the three simple examples considered in this work.

One could wonder whether relaxing the OPERA observation of a constant velocity in the 10-100 GeV energy range could make the strong constraints from neutrino decay compatible with the value $\eta_0=4.7\times 10^{-5}$ measured at an energy of $E=17\,$GeV (mean energy in the OPERA experiment).\footnote{Hereafter in this section we will consider the second example of matrix element which is the most favorable for the superluminal interpretation of the OPERA results.}
Studying the terminal energy $\tilde{E}_n$ or $\tilde{E}_{-\alpha}$ (which is always an upper bound of the average final energy $E_f$) from Eqs.~(\ref{En-km}), (\ref{E-alpha-km}), one sees that  its maximum value is $\tilde{E}_n=20.1\,$GeV. This is however not enough to explain the arrival of much higher energetic neutrinos, so that one would still need to consider a matrix element with much stronger suppression than those considered in this work. Similarly, analyzing the decay length $\ell$ in units of the distance $L$ between CERN and Gran Sasso, we get from Eqs.~(\ref{Gamman-km}) and~(\ref{Gamma-alpha-km})
\begin{equation}
 \dfrac{\ell}{L}\equiv \dfrac{\tilde{\Gamma} _n^{-1}  }{L}=r\, \dfrac{(1+n)^3}{\xi _n}\left(\frac{17\, \mbox{GeV}}{E} \right)^{3n},  \, \mbox{when} \; \; \epsilon (|\vec p|)=\left(\frac{|\vec p|}{\Lambda}\right)^n
 \label{lengthOPERAn}
\end{equation}
and
\begin{equation}
 \dfrac{\ell}{L}\equiv \dfrac{\tilde{\Gamma} _{-\alpha}^{-1}  }{L}=r\, \dfrac{(1-\alpha)^3}{\xi _{-\alpha}}\left(\frac{17\, \mbox{GeV}}{E} \right)^{-3\alpha}, \, \mbox{when} \; \; \epsilon (|\vec p|)=\left(\frac{\lambda}{|\vec p|}\right)^{\alpha}
  \label{lengthOPERAa}
\end{equation}
where
\begin{equation}
r\equiv \dfrac{1.7\times 10^{-5}}{\left[ v(17\, \mbox{GeV}) -1\right]^3}\left(\frac{\text{GeV}}{E}\right)^5\left(\frac{\text{km}}{L}\right)
\end{equation}
and $v(17\, \mbox{GeV})-1=2.35 \times 10^{-5}$ is the velocity at $17$ GeV required for the superluminal interpretation of the results reported by the OPERA collaboration. Eqs.~(\ref{lengthOPERAn}) and~(\ref{lengthOPERAa}) tell us that the decay length is much shorter than $L$ when the energy is  significantly larger than $20$ GeV. This fact can be seen from the dimensionless factor $r$ which is very small for these energies. For example, for $E=50$ GeV, that is roughly a half of the maximum energy detected by OPERA, $r=5.7\times 10^{-3}$ and the maximum value (with respect to $n$ and $\alpha$) of the decay length at this energy is $l=2.8\times 10^{-2}\cdot L$, much shorter than the distance between CERN and Gran Sasso.

\begin{figure}
\centerline{\includegraphics[scale=0.8]{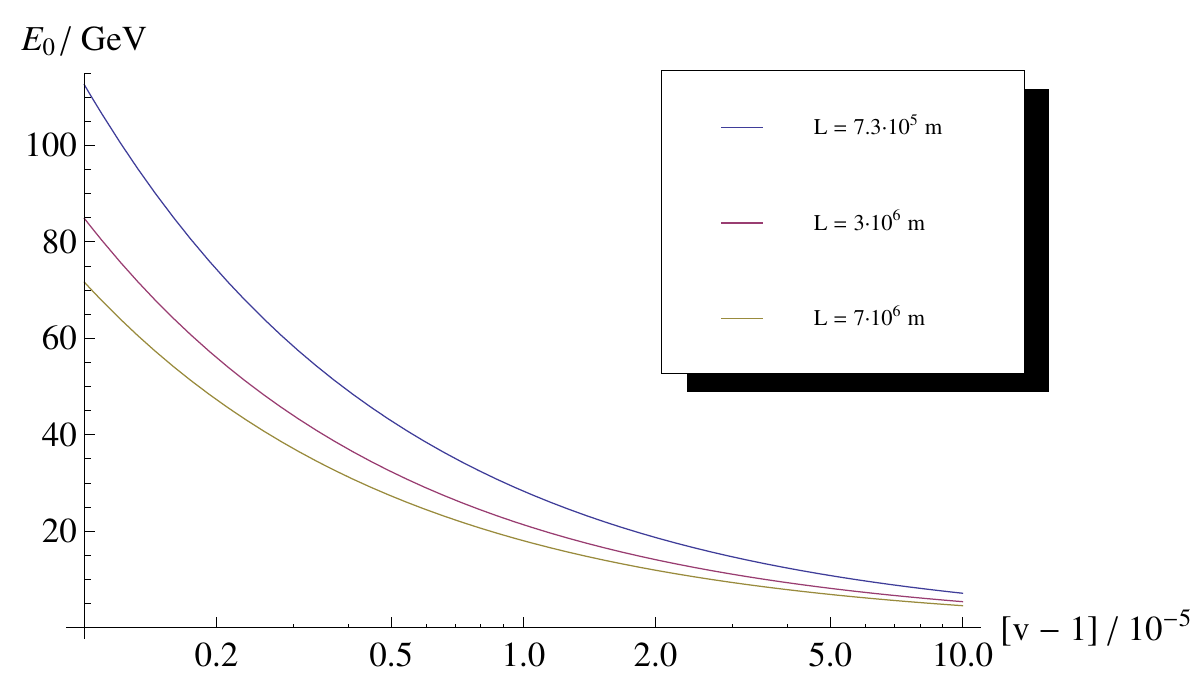}}\hspace{-1cm}
\caption{Value of the terminal energy $E_0$ (approximate final energy of a neutrino with initial energy higher than $E_0$) for different distances of propagation, as a function of the degree of superluminality in the neutrino speed.}
\label{fig:sensitivity}
\end{figure}

The present analysis gives us an idea of the difficulties one finds to accommodate in a theoretical framework the results announced recently by the OPERA collaboration. In fact, the absence of energy loss in neutrino propagation is a more sensitive way to detect  superluminal speeds. Taking our result for the second matrix element in the case of a momentum independent speed of propagation, the observation of the arrival of neutrinos of energies higher than 60\,GeV after a propagation of $L=730\,$km, gives a bound on the possible value of $\eta_0$ through Eq.~(\ref{bounds}),
\be
\eta_0 \leq 5.8 \times 10^{-6},
\ee
which, according to Eq.~(\ref{t0-ns}), corresponds to a difference in the time arrival of neutrinos with respect to luminal speeds of $\Delta t_0 \leq 7.0\,$ns. This means that the non-observation of pair production in OPERA is incompatible (under the assumptions we are considering) with its time-of-flight measurement of $\sim 60$\,ns, and that this non-observation is sensitive to tiny time differences, beyond the present precision of OPERA time measurements.

This conclusion is of interest for any experiment trying to repeat the OPERA measurements. Superluminal speeds can be better detected through a deformation in the spectrum (of course, a direct measurement of the time-of-flight is also advisable, since it does not contain any assumption on the pair production process). In order to appreciate this deformation, we need to send neutrinos with energies higher than $E_0$, which depends on the distance of propagation and the degree of superluminality through Eq.~(\ref{bounds}) (we consider the case of a momentum independent speed as an illustrative case). Therefore, an hypothetical experiment should consider both the neutrino energies of the beam and the distance of propagation in the sensitivity to the degree of superluminality. This sensitivity is shown in Fig.~\ref{fig:sensitivity} in terms of $(v-1)$, the difference between neutrino and photon speeds, using our calculations with the second matrix element (the better motivated theoretically). Fig.~\ref{fig:sensitivity} shows that, in order to be sensitive to $(v-1)$ of the order of $10^{-6}$, we should use neutrino beams of energies higher than 110\,GeV if $L\simeq 730\,$km,\footnote{In 2007 the MINOS collaboration reported a superluminal propagation speed $v-1=(5.1\pm 2.9) \times 10^{-5}$ (at $68\%$ C.L.) for a (mostly muon) neutrino beam with an average energy of 3 GeV propagating 735 km between Fermilab and the Soudan mine~\cite{Adamson:2007zzb}. It is expected that the MINOS team will report new results at higher precision in the near future. The OPERA, ICARUS, BOREXINO and LVD experiments at the Gran Sasso underground laboratory will also provide a more precise measurement of the neutrino speed.} while it would suffice to use neutrino beams of energies higher than 70\,GeV if $L\simeq 7000\,$km (the distance between CERN and the Soudan underground mine in USA).

\section{Concluding remarks}
\label{SecConcluiding}

In this work we have investigated the dependence of the charged lepton pair emission by superluminal neutrinos $\nu _i \to \nu _i \; e^- \; e^+ $ on the dispersion relation for neutrinos, and on the dynamical matrix element of the process. General expressions for an arbitrary dispersion relation and for various examples of matrix elements have been obtained. On the one hand, for a given dispersion relation, different choices of the matrix element lead to decay rates which differ by factors of order one. On the other hand, for a given matrix element, different choices of the dispersion relation lead to energy-independent, dimensionless factors (i.e., pure numbers) in the decay rates which can differ by one order of magnitude. These are the main new results of the present investigation. Estimates of the sensitivity of different observations of high energy neutrinos to a possible departure from SR kinematics in the neutrino sector have been presented. The dependence of the results on details of the theory (choice of modified dispersion relation and modified matrix element) is an indication that high energy neutrino physics can be a very good laboratory to explore possible deviations from SR.

Concerning the generality of the analysis, the assumptions on which it is based were already pointed out in the introduction: i) Rotational symmetry is preserved. Nevertheless, some works have built theories or models in which rotational invariance is not exact (see, for instance, Ref.~\cite{Cohen:2006ky}). ii) Energy and momentum are conserved in the conventional, additive way. However, during the last decade, there have been investigations which suggest the possibility of modifying these laws. Most of these explorations come from the quantum space-time realm and they deal with either deformations (see Ref.~\cite{AmelinoCamelia:2010pd} for a review)  or violations~(\cite{Mer}) of space-time symmetries. iii) The relevant propagation speed of superluminal particles is the group velocity of their wave packets. Once more, several works have studied other alternatives both in canonical and in non-canonical space-times (\cite{Mignemi:2003ab, Daszkiewicz:2003yr, Ghosh:2007ai}).

While experimental results cannot be confirmed or refuted by theoretical investigations but by new experiments, what this present work shows is that the most likely possibility with respect to the observation reported by the OPERA collaboration is that either the observation of superluminal neutrinos is not confirmed or some of the assumptions indicated in the present work are not valid. In particular, in Doubly Special Relativity scenarios~(\cite{AmelinoCamelia:2010pd}), where the energy-momentum conservation law is modified, forbidden processes in SR are generically forbidden too, as it was firstly pointed out in connection with the OPERA results in Refs.~\cite{Carmona:2011zg,AmelinoCamelia:2011bz}.

Some works have studied the pair production decay width before us. In Refs.~\cite{Cohen:2011hx,Huo:2011ve}, the decay width is computed for the case of constant velocity without specifying, however, what  matrix element or relevant Lagrangian (if any) has been used to obtain the result. We reproduce that result for the first example of matrix element which does not come from a $SU(2)$ gauge invariant underlying field theory. The pair production decay width is also computed in Ref.~\cite{Bez} for the momentum independent velocity case with two choices for a four-fermion interaction Lagrangian which are in fact equivalent to our second and fourth matrix elements.

Despite the fact that we have concentrated on the pair production reaction, some of the techniques developed in the present work, and in particular the \textit{collinear} approximation, can be extended to other processes like neutrino splitting $\nu _i \to \nu _i \; \nu _j \; \bar \nu _j$. In the case of a momentum dependent speed for neutrinos, splitting is a relevant mechanism for energy loss in the propagation of superluminal neutrinos.
Other processes contributing to the energy loss of a propagating superluminal neutrino, such as the production of heavier charged leptons and the contribution of virtual $W^\pm$ can also be treated within the same approximation. The results for these processes, as well as an analysis of the uncertainties in the calculation of the production of very high energy superluminal neutrinos with a general dispersion relation will be presented elsewhere~\cite{CCM}.

\section*{Note added}

While this manuscript was being completed, the OPERA collaboration announced the identification of two sources of error in the
determination of the time of flight of neutrinos from CERN to Gran Sasso. One of them was a faulty connection in the optical fiber cable that brings the external GPS signal to the experiment master clock. That this was most probably the origin of the apparent neutrino superluminality was later confirmed (after this paper had been submitted for publication) by the results of muon measurements presented by LVD~\cite{lvd}, which show that the timing between LVD and OPERA became misaligned from the middle of 2008, around the time that OPERA-1 began, and remained stably misaligned by about 73 nanoseconds until the end of 2011, which is when the fiber problem was identified and eliminated. Almost at the same time, the ICARUS collaboration presented new results from the October-November 2011 campaign of measurements, which allowed a very accurate time-of-flight measurement of neutrinos from CERN to LNGS on an event-to-event basis, collecting seven neutrino events which are compatible with luminal speed~\cite{icarus}. Although new measurements are planned during 2012, the new information indicates that most probably the superluminal signal will go away, in agreement with the conclusions of the analysis presented in Section 6.

\section*{Acknowledgments}
This work is supported by CICYT (grant FPA2009-09638) and DGIID-DGA (grant
2010-E24/2).


\end{document}